\input harvmac

\input epsf
\input tables
\def\figin{\epsfcheck\figin}\def\figins{\epsfcheck\figins}
\def\epsfcheck{\ifx\epsfbox\UnDeFiNeD
\message{(NO epsf.tex, FIGURES WILL BE IGNORED)}
\gdef\figin##1{\vskip2in}\gdef\figins##1{\hskip.5in}
\else\message{(FIGURES WILL BE INCLUDED)}%
\gdef\figin##1{##1}\gdef\figins##1{##1}\fi}
\def\DefWarn#1{}
\def\figinsert{\goodbreak\midinsert}
\def\ifig#1#2#3{\DefWarn#1\xdef#1{fig.~\the\figno}
\writedef{#1\leftbracket fig.\noexpand~\the\figno}%
\figinsert\figin{\centerline{#3}}\medskip\centerline{\vbox{\baselineskip12pt
\advance\hsize by -1truein\noindent\footnotefont{\bf Fig.~\the\figno:} #2}}
\bigskip\endinsert\global\advance\figno by1}

\font\zfont = cmss10 
\def\ZZ{\hbox{\zfont Z\kern-.4emZ}}

%

%

%

%

\def\r{R_{11}}

\def\npb{Nucl.Phys.}
\def\phl{Phys. Lett.}
\def\prd{Phys. Rev.}
\def\bb#1{ hep-th/#1}
\def\two{\bf 2}
\def\three{\bf 3}


\lref\nbpsA{A. Sen, Class.Quant.Grav.17:1251-1256,2000}

\lref\nbpsB{A. Sen, \bb9904207.} \lref\nbpsC{A. Lerda and R.
Russo, \bb9905006.}

\lref\nbpsD{J.H. Schwarz, \bb9908126.}

\lref\rotA{S.P. de Alwis, Phys.Lett. B461 (1999) 329;
hep-th/9905080.}

\lref\rotB{A. Mukherjee and M. Sheikh-Jabbari,
/bb0002257.}

\lref\rotC{P. Wang and R. Yue,
\bb9912213.}

\lref\rotD{B. Chen, H. Itoyama, T. Matsuo, and K.
Murakami, /bb9910263.}

\lref\rotE{R. Blumenhagen, L. Gorlich, B.
Kors, \npb B569:209-228,2000; \bb9908130.}\lref\rotF{B.S. Acharya,
J.M. Figueroa-O'Farrill, B. Spence, JHEP 9804:012,1998;
\bb9803260.}\lref\rotG{A. V. Morosov, \phl B433:291-300,1998;
\bb9803110.}\lref\rotH{N. Ohta and P.K. Townsend, \phl
B418:77-84,1998; \bb9710129.}\lref\rotI{M.M. Sheikh Jabbari, \phl
B420:279-284, 1998; \bb9710121.}\lref\rotJ{P.K. Townsend, \npb
Proc.Suppl.67:88-92,1998; \bb9708074.}\lref\rotK{A. Hashimoto and
W. Taylor, IV, \npb B503:193-219,1997;
hep-th/9703217.}\lref\rotL{N. Hambli,\prd D56:2369-2377,1997;
\bb9703179.}\lref\rotM{J.C. Breckenridge,  G. Michaud, and R.C.
Myers, \prd D56:5172-5178,1997; \bb9703041.}\lref\rotN{K. Behrndt
and M. Cvetic, \prd D56:1188-1193,1997; \bb9702205.}

\lref\barbon{J.L.F. Barbon, \phl B402:59-63,1997; \bb9703051.}

\lref\vafa{C. Vafa, {\it Gas of D-Branes and Hagedorn Density of BPS States},
hep-th/9511088, Nucl.Phys. B463 (1996) 415-419.}

\lref\braneReview{D. Kutasov and E. Giveon, Rev.Mod.Phys. 71
(1999) 983-1084 , hep-th/9802067} 

\lref\cw{S. Coleman and  E.
Weinberg, Phys.Rev.D7:1888-1910,1973} 

\lref\books{Books on string
theory such as M. Green, J. Schwarz, and  E. Witten Cambridge, Uk:
Univ. Pr. ( 1987) and J. Polchinski, Cambridge, Uk: Univ. Pr.
(1998)}

 \lref\wb{J. Wess and J. Baggar, Princeton University Press
1992.} 

\lref\juanThesis{J.M. Maldacena, Thesis Princeton University,
``Black holes in string theory,''
hep-th/9607235.
}

\lref\dkps{
M.~R.~Douglas, D.~Kabat, P.~Pouliot and S.~H.~Shenker,
``D-branes and short distances in string theory,''
Nucl.\ Phys.\ {\bf B485}, 85 (1997)
[hep-th/9608024].
}

\lref\brodieHanany{J.~H.~Brodie and A.~Hanany,
``Type IIA superstrings, chiral symmetry, and N = 1 4D gauge theory  dualities,''
Nucl.\ Phys.\  {\bf B506}, 157 (1997)
[hep-th/9704043].
}

\lref\dg{S. Dimopoulos
and H. Georgi, Nucl.Phys.B193:150,1981.} 

\lref\bdl{M. Berkooz, M.
Douglas, and R. Leigh}

\lref\mukhi{S. Mukhi and N. Suryanarayana
hep-th/0003219.} 

\lref\mukhiB{S. Mukhi, N.V. Suryanarayana and D.
Tong, \bb0001066.} 

\lref\evans{N. Evans and M. Schwetz, \npb
B522:69-81, 1998, hep-th/9708122.} 

\lref\senZwiebach{A. Sen and B.
Zwiebach, hep-th/0002211.}

\lref\watiReview{W. Taylor, Trieste
Lectures, hep-th/9801182.}

\lref\ooguri{J. de Boer, K. Hori, H.
Ooguri, and Y. Oz, \npb B522:20-68,1998;
hep-th/9801060.}

\lref\sw{N. Seiberg and E.
Witten, JHEP 9909:032, 1999; hep-th/9908142.}

\lref\add{N.
Arkani-Hamed, S. Dimoloulos, and G. Dvali, Phys.Lett.B429:263-272,
1998;hep-ph/9803315.}\lref\EdThreeD{E. Witten,
hep-ph/0002297.}

\lref\horava{M. Fabinger and P. Horava,
hep-th/0002073.}

\lref\pp{J. Polchinski and P. Pouliot,
Phys.Rev.D56:6601-6606,1997; hep-th/9704029.}

\lref\rs{
L.~Randall and R.~Sundrum,
``Out of this world supersymmetry breaking,''
Nucl.\ Phys.\ {\bf B557}, 79 (1999)
[hep-th/9810155].}

\lref\peskin{
E.~A.~Mirabelli and M.~E.~Peskin,
``Transmission of supersymmetry breaking from a 4-dimensional boundary,''
Phys.\ Rev.\ D {\bf 58}, 065002 (1998)
[hep-th/9712214].}

\lref\how{
P.~Horava and E.~Witten,
``Heterotic and type I string dynamics from eleven dimensions,''
Nucl.\ Phys.\ {\bf B460}, 506 (1996)
[hep-th/9510209].}

\lref\antoniadis{
I.~Antoniadis, E.~Kiritsis and T.~N.~Tomaras,
``A D-brane alternative to unification,''
Phys.\ Lett.\ {\bf B486}, 186 (2000)
[hep-ph/0004214].}

\lref\fernando{
G.~Aldazabal, L.~E.~Ibanez, F.~Quevedo and A.~M.~Uranga,
``D-branes at singularities: A bottom-up approach to the string 
 embedding of the standard model,''
JHEP{\bf 0008}, 002 (2000)
[hep-th/0005067].}

\lref\brodie{
J.~H.~Brodie,
``Fractional branes, confinement, and dynamically generated  superpotentials,''
Nucl.\ Phys.\ {\bf B532}, 137 (1998)
[hep-th/9803140].}

\lref\schmaltz{
M.~Schmaltz and W.~Skiba,
Phys.\ Rev.\ D {\bf 62}, 095005 (2000)
[hep-ph/0001172].}

\lref\luty{
Z.~Chacko and M.~A.~Luty,
``Radion mediated supersymmetry breaking,''
hep-ph/0008103.}

\lref\rsTwo{
L.~Randall and R.~Sundrum,
``An alternative to compactification,''
Phys.\ Rev.\ Lett.\ {\bf 83}, 4690 (1999)
[hep-th/9906064].
}

\lref\rsThree{
L.~Randall and R.~Sundrum,
``A large mass hierarchy from a small extra dimension,''
Phys.\ Rev.\ Lett.\ {\bf 83}, 3370 (1999)
[hep-ph/9905221].
}

\lref\joe{
J.~Polchinski,
``TASI lectures on D-branes,''
hep-th/9611050.
}

\def\vect{\vec}
\def\vac{|0>}

\def\daggar{\dag}
\def\ad{a^{\daggar}}

\def\bd{b^{\daggar}}

\def\mF{(-1)^F}
\def\CW{Coleman-Weinberg}
\def\g{g^2_{YM}}

\def\cpl{{1\over \g }}

\def\ra{\rightarrow}
\def\gg{g_{YM}}
\def\ls{\sqrt{\alpha'}}
\def\r{R}
\def\l{L_6}

\def\str{Str}
\def\dd{\partial}
\def\hf{{1\over 2}}


\def\LongTitle#1#2#3#4#5{\nopagenumbers\abstractfont
\hsize=\hstitle\rightline{#1}
\hsize=\hstitle\rightline{#2}
\hsize=\hstitle\rightline{#3}
\vskip 0.5in\centerline{\titlefont #4} \centerline{\titlefont #5}
\abstractfont\vskip .3in\pageno=0}

\LongTitle{hep-th/0101115}{SLAC-PUB-8706}{}
{On Mediating Supersymmetry Breaking}
{in D-brane Models}

\centerline{\it John H. Brodie}
\centerline{\it Stanford Linear Accelerator Center}
\centerline{\it Stanford University}
\centerline{\it Stanford, CA 94305, USA}
%

%


\vskip 0.3in
\centerline{\bf Abstract}
\bigskip

We consider the 3+1 visible sector to live on a 
Hanany-Witten D-brane construction in type IIA string theory.
The messenger sector consists of stretched strings from the 
visible brane to a hidden D6-brane in the extra spatial dimensions.
In the open string channel supersymmetry is broken by gauge mediation while 
in the closed string channel supersymmetry is broken by 
gravity mediation. Hence, we call this kind of mediation ``string 
mediation''. We propose an extension of the Dimopoulos-Georgi theorem 
to brane models: only detached probe branes can break 
supersymmetry without generating a tachyon. Fermion masses are generated 
at one loop 
if the branes break a sufficient amount of the ten dimensional Lorentz group 
while scalar potentials are generated 
if there is a force between the visible brane 
and the hidden brane. Scalars can be lifted at two loops through 
a combination of brane bending and brane forces. We find a large 
class of stable non-supersymmetric brane configurations of ten 
dimensinoal string theory.

\Date{January 2001}

\def\sugra{
Since the long distance limit of the close string theory is
supergravity we can calculate the long distance force between
branes using only the brane action and supergravity. We will
explain how to do that now (see also for example
\pp\dkps\watiReview). In the configuration of branes with 4ND
boundary conditions, where N is for Neuman and D is for Dirichlet,
the velocity independent forces cancel between graviton and
dilaton \foot{ In the 0ND supersymmetric limit the graviton and
dilaton cancel against the RR field.}. In the non-supersymmetric
limit where the supergravity forces don't cancel, they appear in
the brane action as potentials for the scalar field. The way this
works is the following: The metric for a p'-brane is
\eqn\metric{ds^2 = f(r)^{-1/2} dx_{||}^2 + f(r)^{1/2} dx_{perp}^2}
where $x_{||}$ are the coordinates parallel to the brane,
$x_{perp}$ are the coordinates perpendicular to the brane, and
\eqn\harmonic{f(r) = 1 + g_s ({\sqrt{\alpha'}\over r})^{7-p'}.}
The dilaton obeys the equation \eqn\dilaton{e^{-2\phi} =
f^{p'-3\over 2}} (see for example \juanThesis\dkps.) The action
for the p-brane is \eqn\pAction{S_p = \int d^{p+1}y e^{-\phi}
\sqrt{\det G(y)}} where $G_{\mu\nu}(y)$ is the pullback of the 10d
p'-brane metric onto the p-brane. \eqn\pullback{G_{\mu\nu}(y) =
h_{IJ}(x) \dd_{\mu}x^I \dd_{\nu} x^J} where $h_{IJ}(x)$ is given
by the line element in \metric\ and $\mu = 0...p$ and $I = 0
...9$.

Plugging \metric\ and \dilaton\ into \pAction\ we find for
parallel p and p'-branes \eqn\potAndBeta{S_p = f(r)^{p'-3\over 4}
f(r)^{-(p+1)\over 4} (1 + f(r) \dd_{\mu} X^i \dd^{\mu} X_i + ...)}
where $i=p+1,...,9$. Expanding out the potential in \potAndBeta\ we
see that
\eqn\pot{V_p(r) = 1 + ({p'-3\over 4})({\sqrt{\alpha'}\over
r})^{7-p'} - ({p+1\over 4})({\sqrt{\alpha'}\over r})^{7-p'} +
\cdots} For $p'-p=4$ the dilaton force cancels the gravitational
force. For Dp parallel to a Dp' brane this is corresponds to the
4ND supersymmetric boundary condition.

Now let us consider a non-supersymmetric configuration of BPS branes at an
angle $\theta$ which will be our main way of breaking supersymmetry as
will be discussed in more detail below.
Because of the form of the metric \metric, directions of the
p-brane parallel to the p'-brane increase the exponent of $f$ in
the potential \pot\ by ${1\over 4}$ while directions of the
p-brane perpendicular to the p'-brane decrease it by ${1\over 4}$.
For a brane at an angle $\theta$ the formula for the potential is
then \eqn\potAngle{V_p = f^{p'-3\over 4} f^{-p\over 4}(\sin^2 \theta
f^{1/2} + \cos^2 \theta f^{-1/2})^{1\over 2}} For the
case $p'-p=2$ and $\theta =0$, the potential \potAngle\ vanishes.
This is the 4ND condition for a Dp perpendicular to a Dp'.
However for very small $\theta$ and
$r>>\sqrt{\alpha'}$ \potAngle\ becomes \eqn\smallAnglePot{V_p =
M_s^{p+1}({1\over g_s} + \theta^2 ({\ls \over r})^{7-p'} +\cdots)} If we set
$\phi = M_s^2 r$, then \smallAnglePot\ is the potential that a
scalar field, $\phi$, on the p-brane experiences.}

\def\spectrum{
The first thing we have to understand about the brane
configuration is the spectrum of states on the stretched strings
between a p-branes and a p'-brane. The spectrum is well known and
is given in terms of the string partition function \books. If the
boundary conditions are NN (Neuman on the p-brane -Neuman on the
p'-brane) in directions $\mu = 0,\dots ,p$ or DD in directions $I
= p'+1,\dots  ,9$ , then in the NS sector there can be only
periodic bosons $a^{\mu}_{-n}$ and $a^{I}_{-n}$ and anti-periodic
fermions $b^{\mu}_{-{n\over 2}}$ and $b^{I}_{-{n\over 2}}$
oscillating in those directions where $n=0\cdots\infty$ for
periodic oscillators and $n=1,\cdots ,\infty$ in the case of
anti-periodic oscillators. The ND directions $i=p+1,\cdots ,p'$
have antiperiodic bosons $a^{i}_{-n\over 2}$ and $a^{i}_{-n\over
2}$ and periodic fermions $b^{i}_{-{n}}$ and $b^{i}_{-{n}}$.
Bosons and fermions get quantized by the usual rules
\eqn\qz{\eqalign{[a_n,\ad_m] & = \delta_{n-m}\cr \{ b_n,\bd_m\} &
= \delta_{n-m} \cr }} Periodic fermions allow for fermionic zero
modes \eqn\fzm{\{ b^{i}_0,b^j_0\} = \delta^{ij}} Quantizing the
zero modes gives rise to $2^{p'-p\over 2}$ states degenerate with
the vacuum. The 2d vacuum energy gets shifted in the NS sector by
$-1/24$ by the periodic bosons, $-1/48$ by the antiperiodic
fermions, $+1/24$ by the periodic fermions, and $+1/48$ by the
anti-periodic bosons. The total vacuum shift in the NS sector is
then \eqn\shift{E_{NS}^0 = (8-\nu)(-1/24 - 1/48) + \nu (1/24 +
1/48)} where $\nu =p' - p$ is the number of ND directions. If $\nu
< 4$ then there is a tachyon in the NS sector. Defining
\eqn\parts{\eqalign{ f_1(q) & = q^{1\over 12} (1-q^{2n}) \cr
f_2(q) & = q^{-1\over 24} (1+q^{2n-1}) \cr f_3(q) & = q^{-1\over
24} (1-q^{2n-1}) \cr f_4(q) & = q^{1\over 12} (1+q^{2n}) \cr }}
The NS partition function is \eqn\nsp{Z_{NS} = 2^{\nu\over
2}{f_2^{8-\nu} f_4^{\nu} \over f_1^{8-\nu} f_3^{\nu}}}

On the Ramond string, the vacuum energy vanishes (due to
supersymmetry). The NN and DD sectors have periodic bosons and
periodic fermions while the ND sector has anti-periodic bosons and
anti-periodic fermions. In the R sector there is $2^{8-p-p'\over
2}$ states from quantizing the fermionic zero modes in the light
cone gauge. The partition function for states on the Ramond string
is \eqn\nsp{Z_{R} = 2^{8-\nu\over 2}{f_4^{8-\nu} f_2^{\nu} \over
f_1^{8-\nu} f_3^{\nu}}}

According to the GSO projection, one must project out all states
that satisfy \eqn\gso{\mF = +1} For space-time supersymmetry we
must have \eqn\ssy{Z_{NS} - Z_R =0} }

\def\probe{

In this section we will explain how supersymmetry breaking can be
communicated to a visible p-brane from a rotated probe p'-brane
located a distance $\r$ away via fundamental strings. The
combination of the p-brane and the p'-brane breaks supersymmetry
globally in the ten dimensional space-time. The strings that
stretch between the p-brane and the p'-brane which we call p-p'
strings are the heavy messenger fields that communicate the
supersymmetry breaking to the light fields on the visible brane.
Supersymmetry breaking arises due to the quantum mechanical zero
point fluctuations of the heavy messenger fields. The zero point
energy induces a vacuum energy and masses for the supersymmetric
massless fields in the visible sector which couple to the
non-supersymmetric messenger fields through the gauge fields.
Let's do an example where $p=4$ and $p'=6$. The messengers will be
supersymmetric before we turn on the rotation. Later we will
consider rotations. The 4-4 strings are the visible sector. The
6-6 strings are hidden on the probe brane and the 4-6 and 6-4
strings are the supersymmetric messenger fields. We understand
that the messenger fields are supersymmetric in the following way:
According to the equation for summing up the zero-point energies
on the NS string (see Appendix C) \eqn\shift{E_{NS}^0 =
(8-\nu)(-1/24 - 1/48) + \nu (1/24 + 1/48)} the vacuum energy for
the NS string is zero since $\nu$ the number of ND boundary
conditions equals four. The vacuum energy of the Ramond sector is
always zero by supersymmetry. So the vacuum energies for the NS
and the R sector match. Moreover, there are fermionic zero modes
that come from the four ND and DN boundary conditions in the NS
sector and from four NN and DD boundary conditions in the R
sector. Using these four fermionic zero modes we can build
$2^{4/2} = 4$ states including the vacuum state, half of which are
killed by the GSO projection. Including the oppositely oriented
fundamental string there are a total of 4 states from the NS
sector and 4 from the R sector. All of these states are fermionic
in the 10d space-time, but because the D4-brane and D6-brane break
the $SO(9,1)$ space -time Lorentz group down to $SO(1,3)\times
SO(2)_J \times SO(3)_R$ from the 3+1 point of view the R states
are world-volume fermions charged under the 3+1 Lorentz group as
well as the $SO(2)_J$ R-symmetry but are scalars under the
$SO(3)_R$. We will call these states $\psi_q$ and $\psi_{\tilde
q}$. The states that come from the NS-string are fermions under
the $SO(3)_R$ R-symmetry and scalars under the world-volume
Lorentz group. We will call these fields $q$ and $\tilde q$.
Together these fields make up a $N=2$ hypermultiplet, $Q$ and
$\tilde Q$. Therefore, the messengers are supersymmetric, and
there is no supersymmetry to communicate. Now let us consider a
relative rotation of the D4-D6 system: Take the D4-brane to extend
in 01236 and the D6-brane to extend in 0123789. Let us now rotate
the D6-brane in the 67-plane by an angle $\theta$. If $\theta =
{\pi\over 2}$, then D6 extends in 0123689. What happens to the
fundamental string during the rotation? Some periodic fermions on
the worldsheet  become anti-periodic,  and hence there won't be as
many fermionic zero modes are before. Moreover, since the vacuum
energy on the string is the sum of the zero point energies of all
the modes on the string (periodic bosons and fermions,
anti-periodic bosons and fermions) the rotation will change the
vacuum energy as well. Percisely, for $\theta = {\pi\over 2}$ $\nu
=2$, equation \shift\ equals $-{1\over 4\alpha'}$. Therefore the
mass of the NS sector is tachyonic. During the rotation, the mass
of $q$ went up by ${\theta\over 2\pi\alpha'}$ while the mass of
$\tilde q$ was lowered by $-{\theta\over 2\pi\alpha'}$ . The mass
of the states $\psi_q$ did not change since the rotation in the
67-plane did not effect the periodic fermions in 2345. Moreover,
it is a general property of these rotations that all the massive
and massless multiplets preserve \eqn\strace{\str M^2  = 0} as is
true of softly broken supersymmetric theories as we saw above in
section 2.3 . }


\def\intersection{All this leads us to conclude that
the low energy theory on the intersection of $N_c$ visible
D4-branes with $N_f$ hidden D6-branes  is a 3+1 $SU(N_c)\times
U(1)$ $N=2$ gauge theory \foot{The D4-brane is of course 4+1
dimensional. We will compactify it later.} with an massless
adjoint hypermultiplet $A$ and $B$ and as we saw above heavy $N_f$
fundamental hypermultiplets playing the role of the messenger
fields, \eqn\hyper{\eqalign{Q & = q + \theta \psi_q + \theta\theta
F_q \cr \tilde Q & = \tilde q + \theta \tilde \psi_{q} +
\theta\theta \tilde F_q}.} where $q$ has spin 0 and
$\psi_q$ has spin ${1\over 2}$. There are many massive modes
corresponding to the excited 4-6 open string states. These are
massive fundamental fields with spin $J = 1, {3\over 2}, 2,...$.
Turning on a non-zero angle in the 6-7 direction corresponds to
turning on a FI term. The Lagangian is \eqn\action{\eqalign{{\cal
L}_{N=2} & = \int d^2\theta d^2\bar\theta {1\over\g }
(\Phi_a^{\dag} e^{V_b} \Phi_c f^{abc}) + Q_i^{\dag} e^{V_a
T^a_{ij}} Q_j + \tilde Q_i^{\dag} e^{- V_a T^a_{ij}} \tilde Q_j +
\eta \Tr V \cr & + \int d^2\theta {1\over g^2_{YM}}W^{\alpha}_a
W^a_{\alpha} + \int d^2\theta (mQ\tilde Q+ \lambda \Phi Q\tilde Q
+ \Phi [A,B]) \cr }} $a$ is now an adjoint valued gauge index and
$i$ is a fundamental index. In terms of component fields there is
a potential for the $q$-fields \eqn\pot{V = |m+\lambda\phi |^2
|q|^2 + |m+ \lambda \phi |^2|\tilde q|^2 + \g (|q|^2 -
|\tilde q|^2 + \eta)^2} The adjoint fields $\Phi, A$, and $B$ are
neutral under the diagonal $U(1)$ and therefore do not couple to
the FI term, but the fundamental fields do. From \pot\ we see that
the mass of $q$ is raised by $\eta$ while that of $\tilde q$ is
lowered by $\eta$. Higher spin fields that are in the fundamental
representation will also be split accordingly. Note that massive
higher spin multiplets in the adjoint representation will not be
split by the rotation and so supersymmetry is broken only at the
loop level for the adjoint fields.}

\def\annulus{
The fundamental process we will use for communicating
supersymmetry breaking is the string annulus diagram. In string
theory, the annulus diagram, a two-dimensional world sheet with
one hole and two boundaries, has two interpretations: with time
running radially, it is a close string exchange between two
tensionful branes. With time running vertically, it is an open
string loop. The close string interpretation is more useful when
the branes are separated by a distance
much larger than $\ls$. In this limit, the
supersymmetry breaking is communicated via tree level gravitons
from the hidden brane to the visible brane, and massive string
modes can be neglected. In fact, in this paper all of the ``field theory''
calculations will be done using the p-branes in the supergravity background
of a p'-brane. The open string interpretation is more
useful when the branes are separated by a distance less than $\ls$.
In this limit the supersymmetry breaking is communicated via field
theory loop diagrams of light scalar and fermion
fields that propagate in both the
hidden and the visible brane.}

\def\magnetic{

Until now we have broken supersymmetry by rotating the p'-brane relative
to the p-brane. Here we will show that rotating the probe brane is
equivalent to turning on a flux in the probe brane. For example,
consider $N_c$ 4-branes and tilted $N_f$ 6-branes system at an
angle $\theta$. A higher energy excitation of this system is $N_c$
D4s along 01236, $N_f$ D6s along 0123789 and $k$ D6s along 0123689
where $k$ is related to the angle $\theta$\watiReview. T-dualizing
along the 6-direction one gets $N_c$ 3-branes and $N_f$ 7-branes
and $k$ D5-branes. Letting the system relax back to its ground
state, we find $N_c$ D3-branes and $N_f$ D7s with $k$ units of
magnetic flux along its worldvolume. The flux in the 7+1
worlvolume theory points along 67.
There is then a coupling to the spinor of $SO(4)_R$ in directions
6789 \eqn\spinCoupling{H = F_{MN}\sigma^{MN} \rightarrow 
\dd_{M}\phi_{N}\sigma^{MN}} For these
purposes we can ignore the directions 0123 and focus on 6789 and
then the problem becomes the familiar one of a spin 1/2 particle
in a magnetic field in four dimensions. The spectrum splits: spin
up particle along the B-field have their energy lowered while spin
down particles have their energy raised. Because the fermions and
bosons are no longer degenerate in mass, supersymmetry is broken.
Of course in the string theory models 
what is relevant are the fermions in the R-symmetry group 
which are scalars under the 3+1 Lorentz group and 
the magnetic flux is the flux in the 7+1 flavor fields not the 
3+1 color fields.
The arrow pointing to the right in equation \spinCoupling\ 
indicates that under T-duality 
the flavor flux turns into the gradient of a scalar field. 
Note that the gradient does not have to be constant as was seen in 
section 3.6. $\theta$, $F_{MN}$, $\dd_{M}\phi_{N}$, are all proportional 
to the D-term.}

\def\hwconst{

\ifig\hw{This is a brane configuration for $N=2$ super Yang-Mills}
{\epsfxsize2.0in\epsfbox{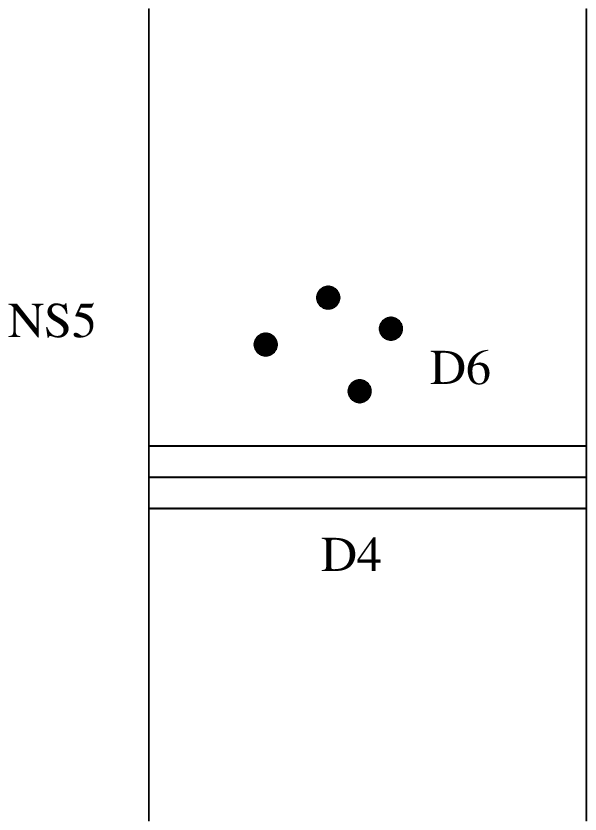}}

\ifig\rot{Rotation of the D6-brane in the 6-7 plane corresponds
to a FI-term in the field theory on the brane. The vertical
dotted line is not a brane; it is just a reference.}
{\epsfxsize2.0in\epsfbox{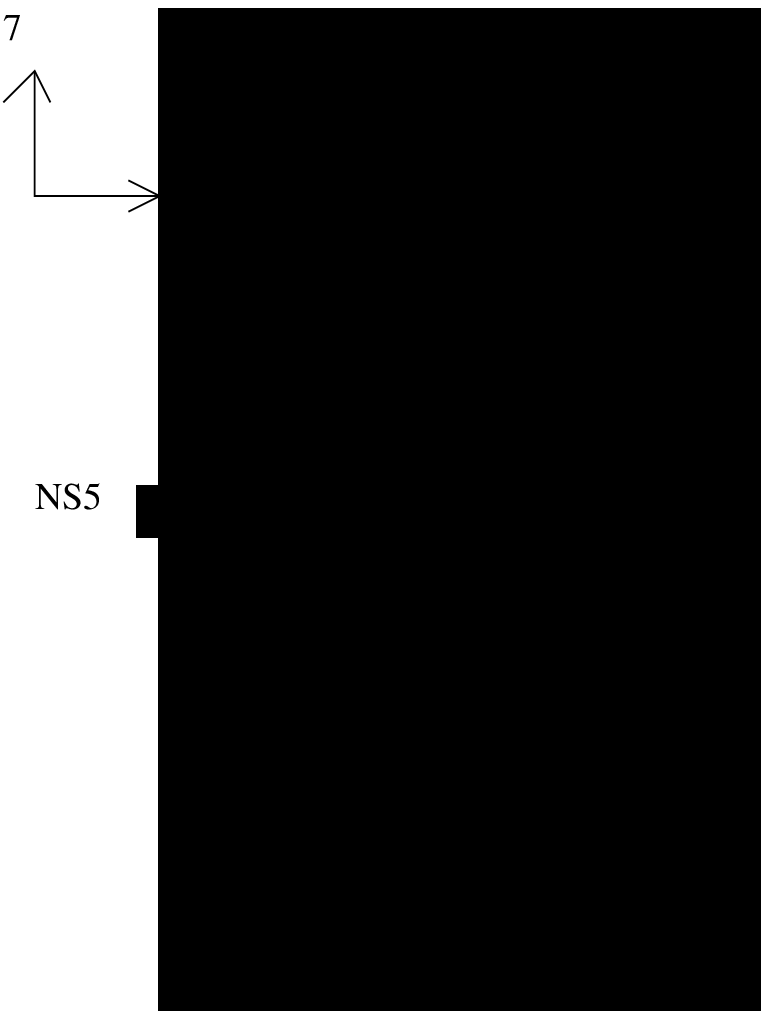}}

Let us consider NS5 branes in directions 012345, D4 01236, and D6
in directions 0123789 (see \hw ). The NS-branes serve to restrict
the motion of the D4-brane to the 45-directions. In addition the
two NS5-branes dimensionally reduce the 5d theory on the D4-brane
to four dimensional massless fields plus Kaluza-Klein modes with
mass ${1\over \l}$. On the worldvolume of the branes, the action
is \action\ with the addition of mass terms for two of the three
complex adjoint scalar fields \eqn\massAdj{W = mA^2 + mB^2.} The
NS5-branes will not participate in the supersymmetry breaking to
lowest order. The $N_f$ D6-branes are the probe branes. The global
symmetries $SO(1,3)\times U(1)_J \times SU(2)_R$ which correspond
to the broken ten dimensional Lorentz symmetry
$SO(1,3)_{0123}\times SO(2)_{45} \times SO(3)_{789}$. The fields
in \action\ transform in the way shown in the table below.
The $N=1$ $U(1)_{45+89}$
symmetry is the sum of the $U(1)_{45}$ and the $U(1)_{89}$ and the
gaugino $\lambda$ has been normalized such that it has charge 1
under the $N=1$ R-symmetry. This is the R-symmetry that survived after 
the adjoint chiral multiplet gets a mass.

\ifig\split{The spectrum on the left is the supersymmetric
spectrum where the fermions $\psi_q$ are degenerate with the
bosons $q$ and $\tilde q$. The spectrum on the right corresponds
to turning on an angle in the branes and to turning on an FI-term
in the field theory. $q$ gets a mass that raises it up while
$\tilde q$ gets lowered in mass. Note that the sum of the masses
doesn't change. Other multiplets charged under the fundamental
representation are split in a similar way.
Notice that the spectrum 
looks like a spinor in a magnetic field. This is no coincidence 
since the scalar fields $q$ are in fact spinor in the R-symmetry group. 
Moreover, the D-term can be thought of as a magnetic field in those directions.
In the brane picture this is clear: rotation T-dualized into magnetic flux.
}
{\epsfxsize2.0in\epsfbox{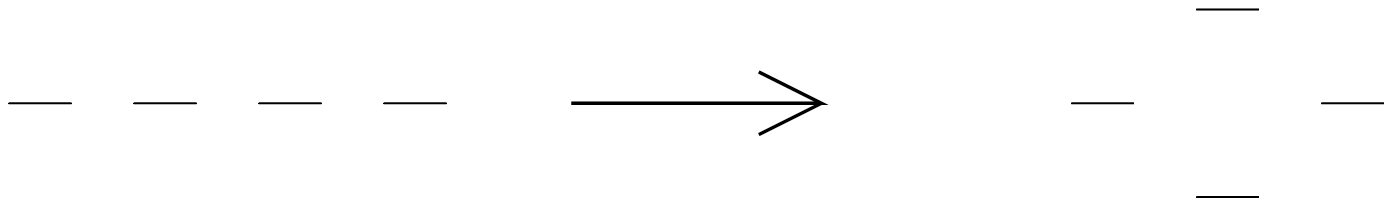}}}

\def\evansParameters{

Rotating the 6-brane in directions 67, 68, or 69 corresponds to a
real FI D and complex $F_{\phi}$-term. The number of parameters is
equal to the number of rotations of a line in four dimensions:
$SO(4)/SO(3)$ as is explained in \evans. One can also rotate the
D6-brane in 789 and 45. This is the rotation of a 3-plane in 5
spatial dimensions given by $SO(5)/SO(3)\times SO(2)$ which has 6
generators. Notice that we are moding out $SO(5)$ precisely  
by the R-symmetry group. 
One can identify the spurions are having charge under
the R-symmetry group; they form a 6 dimensional $(2,3)$ representation 
under $SO(2)\times SO(3)$. The fact that there 
are only two gauge invariant operators that one can write 
for a representation $(2,3)$ maps nicely into the fact that there are only 
two invariant rotations of the D6-branes relative to the 
D4-branes. We can identify these rotations $\alpha$ and $\theta$ with the Yukawa
coupling $\lambda$ in the superpotential \action\ as well as 
non-holomorphic terms
\eqn\nonholSup{W = \sin \alpha((1+\cos\theta )\Phi +
(1-\cos\theta)\Phi^{\dag})Q\tilde Q.} Other 
operators correspond to field redefinitions of 
\nonholSup\ such as 
\eqn\Redef{{\cal L} = \xi \int d\theta_{\alpha}W^{\alpha}Q
\tilde Q = \xi \lambda q \tilde \psi_q.} Rotating the D6-brane
$\theta = \pi$ in the 75 plane as can be seen in \nonholSup\ takes the usual
superpotential coupling \eqn\holoToHolo{\Phi Q\tilde Q \ra
\Phi^{\dag} Q\tilde Q.} In terms of components this eliminates the
Yukawa coupling of the adjoint fermion to the squark and quark
fields. In fact, the configuration with $\theta = \pi$ was
investigated in \mukhi\mukhiB\ where they found it to have exactly this
interpretation in terms of components. 

There are
also the rotation of the two NS 5-branes with respect to each other
as was discussed in \evans.  
$SO(5)/SO(2)\times SO(3)$ again have 6 generators. This we can identify
with the mass term for the adjoint $m$ \barbon, a charge $(-1,1)$
under the R-symmetry group $U(1)_{45}\times U(1)_{89}$.
The operator
\eqn\massOp{W = m\Phi^2}
breaks the R-symmetry 
to a single diagonal $U(1)$.
Geometrically this will be a linear combination 
of $U(1)_{45}$ and $U(1)_{89}$.
The parameter $\zeta$ in 
the operator
\eqn\thirdOp{{\cal L} = \zeta \int d\theta_{\alpha}W^{\alpha}\Phi 
= \zeta \lambda \psi + \zeta D \phi}
has charge $(-1,0)$. Finally, there is the parameter
$F_{\tau}$ which in the $N=2$ action is correlated with the 
parameter $\mu$ in the operator 
\eqn\nonholoMassTerm{W = \int d^2\bar \theta \mu \Phi^{\dag}\Phi
= \mu F_{\phi}^* \phi}
both of which have charge $(-1,-1)$ under the R-symmetry group.
Altogether these spurions make a ${\bf 3_{-}}$ under 
the $U(1)_J\times SU(2)_R$. The ${\bf 3_{+}}$ comes from 
the hermitian conjugate operators in the 
superfield Lagrangian.
 One can write the non-holomorphic mass term in the superpotential
\nonholoMassTerm\ in the more geometrically suggestive way
\eqn\WnonHolo{W =  \sin\beta((1+\cos \omega)\Phi + (1-\cos
\omega)\Phi^{\dag })\Phi}
where the mass scale is set by the separation between the 
branes $1/L_6$.
The angles $\beta$ and $\omega$ correspond to the 
two invariant operators one can form from the six dimensional 
representation $(2,3)$ of $SO(2)\times SO(3)$.
Interestingly, for $\omega = 0$ there is a point with 
a single massless real adjoint scalar while the other adjoint scalar is 
massive as is shown in figure 12. At this point both adjoint fermions have the same 
mass preserving a $U(1)$ global symmetry that can be seen 
geometrically as a subgroup of the $SO(3)_{789}$.
In terms of operators the real scalar can be lifted by turning 
on $\zeta$ in equation \thirdOp. It is clear that $U(1)_{89}$ will be preserved 
since $\zeta$ is not charged under it.
It is easy to show that all of these rotations satisfy 
$Str M^2 = 0$.

 One can consider the rotation of the 2 NS
5-branes from 45 into 6. This is the rotation of a plane in 3
dimensions: $SO(3)/SO(2)$ which has 2 generators. We don't have
any interpretation of these to offer. Moving the D6-brane in the
45 direction corresponds to adding a supersymmetric mass term
\eqn\WmassQ{W = \int d^2\theta m Q\tilde Q} .}

\def\trig{

In N=2 SQED, turning on both a D-term, $\eta$, and a mass term,
$m$, forbids a supersymmetric configuration \wb. For $m<\eta $,
the theory is forced onto the Higgs branch as one can see from the
potential \eqn\Hpot{V = m^2|q|^2 + m^2|\tilde q|^2 + \g (|q|^2 -
|\tilde q|^2 + \eta)^2 } since $\tilde q$ is tachyonic. Since D
and F are a $\bf 3$ of the $SU(2)_R$ turning on the D-term breaks
this symmetry to $U(1)_R$. The mass has charge 2 under the $U(1)_J$
whose vacuum expectation value breaks the symmetry completely. For
non-zero FI-term and zero supersymmetric mass term there is a
supersymmetric configuration on the Higgs branch but the Coulomb
branch is lifted. One can compare the top of the potential for
$\tilde q$ with the bottom of the potential and see that the brane
theory reproduces the same difference in vacuum energy as the
field theory. To demonstrate this consider that, in the brane
theory, the difference in vacuum energy from the unHiggsed to the
Higgsed configuration is given by the difference in the length of
the D4-brane from the long configuration when it stretches between
the two NS5-branes to the shorter configuration where it splits
into 2 pieces which stretch between the NS5 and the rotated
D6-brane. \eqn\length{ E^4 = T_{D4}(L-D)} where
\eqn\tension{T_{D4} = {M_s^5\over g_s}}
 is the tension of the D4-brane,
$L$ is the length of the brane stretched between in NS 5-branes,
and D is the sum of the length of the branes from the NS5 to the
D6. They are related by trigonometry \eqn\trig{D = L\cos\theta =
L(1-{\theta^2\over 2} + ...)} Where $\theta$ is the angle between
the D4 and the D6. Plugging \trig\ into \length\ we find that
\eqn\len{E^4 = {T_{D4}L\theta^2\over 2}} Now $L$ is related to the
Yang-Mills coupling on the D4-brane and $\theta$ is related to the
FI-term as \eqn\parameters{\eqalign{L & = {g_s\over \g M_s} \cr
\theta & = {\g\eta\over M_s^2}\cr }} Inserting \parameters\ and
\tension\ into \len\ we learn that \eqn\energyfour{E^4 =
{M_s^5\over g_s}{g_s\over \g M_s}{\gg^4\eta^2\over M_s^4} =
\g\eta^2 } which agrees with the vacuum energy given by equation
\Hpot.

\ifig\higgs{This minimum length
configuration for the D4-brane is split
between the D6 and the NS5.
For $m = 0$, this is a
supersymmetric Higgs branch in the field theory.}
{\epsfxsize2.0in\epsfbox{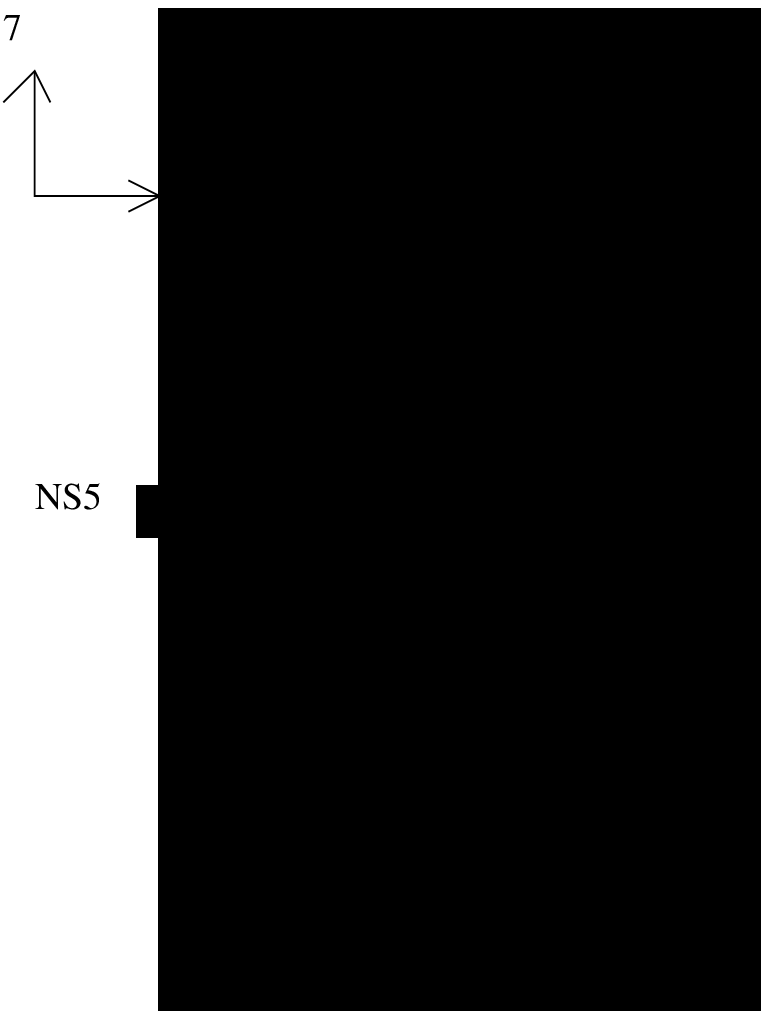}}

\ifig\noSusy{Here the D4-brane has come so close to the D6-brane
that it has split into a D4-brane and an anti-D4-brane. In the
field theory on the brane, supersymmetry and gauge symmetry are
broken. Note that there are two limits one can take here: If
length scales are big compared to the string scale $\ls$, then
there is a first order transition. The branes will jump from one
configuration to the other more energetically favorable
configuration without inducing a tachyon. However, if length
scales are small compared to the string scale, then stretched
strings will become so light that a tachyon is induced. There is
then a second order phase transition from a Coulomb phase into a
Higgsed phase.} {\epsfxsize2.0in\epsfbox{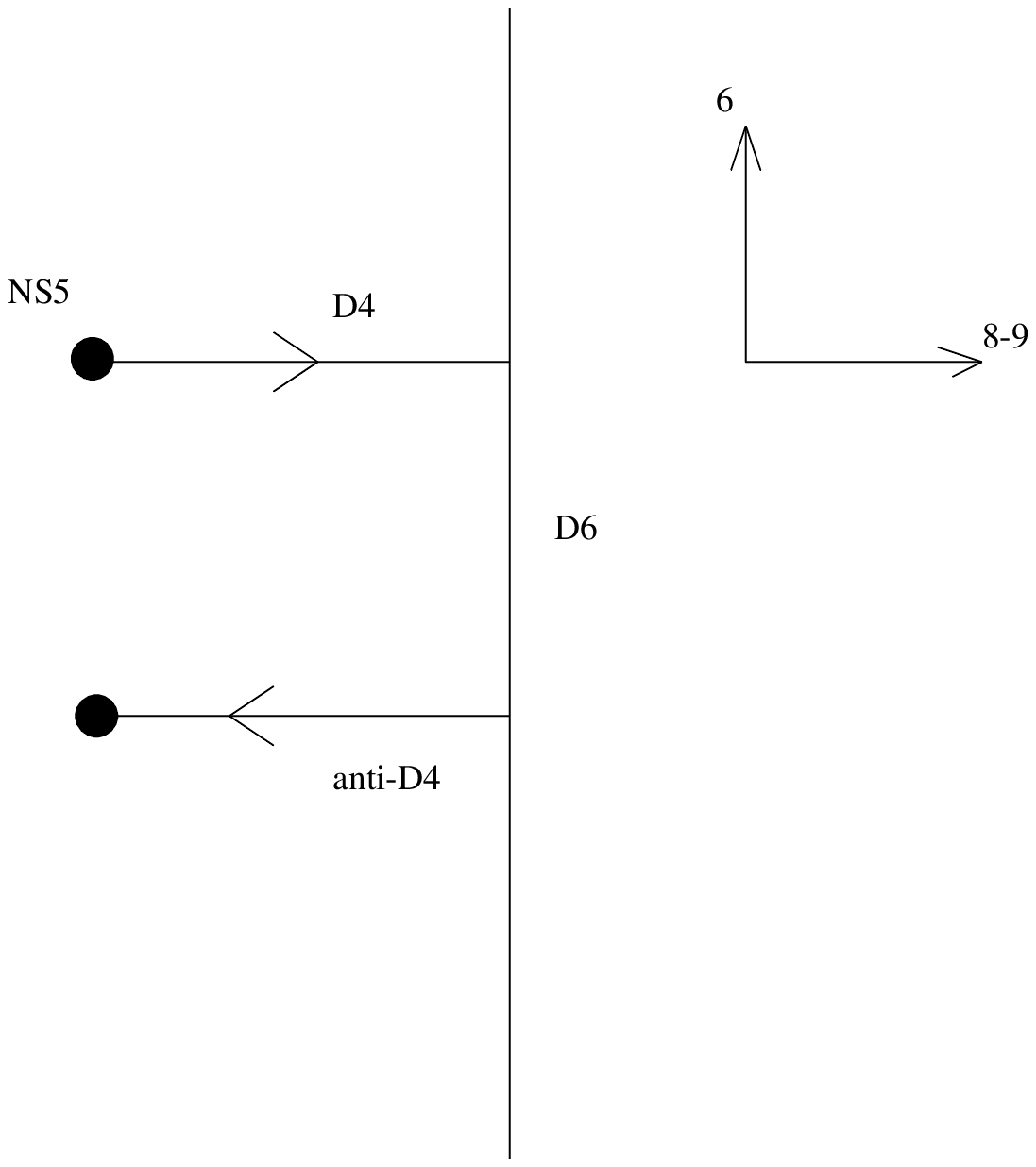}}

}
\def\potentials{

In the open string channel, the strings are charged under the
hidden and visible sector. Quantum mechanically the heavy
messenger fields have a zero point energy which contributes to the
vacuum energy of the theory and can give rise to scalar
potentials. Let us review some facts about calculating such
potentials in quantum field theory using the methods of Coleman
and Weinberg. Supersymmetric theories are special in having flat
potentials since the fermion and boson zero point energies exactly
cancel. Generically, in non-supersymmetric theories potentials are
generated by quantum corrections to the vacuum energy at 1-loop
\cw. \eqn\vac{\eqalign{V = \sum_{\omega} (-1)^F {1\over 2} \omega
& = \sum_I (-1)^F \int d^{D-1} k \sqrt{\vect k^2 + m_I^2} \cr =
\sum_I (-1)^F \int d^Dk \log{(k^2 + m_I^2)} & = \sum_I (-1)^F \int
d^Dk \int {dt\over t} e^{-t(k^2 + m_I^2)} \cr = \sum_I (-1)^F \int
{dt\over t^{D/2+1}} e^{-m_I^2 t} \cr }} In softly broken
supersymmetric field theories with one light multiplet of mass
$M$, these potentials take on the form \eqn\ftpot{V = \delta M^4
\log M^2} where $M$ is the supersymmetric mass and where
\eqn\strace{\str M^2 = \Tr (-1)^F M^2 = 0.}  Here $F$ is the number
of fermions and $M$ is the mass of the multiplet. However, a
theory with an infinite tower of massive fields such as string
theory can yield different potentials after one sums all the
contributions from the individual fields
.\eqn\stPot{\sum_{I=0}^{\infty} \delta M^4 \log (M_I^4).} In fact,
string theory tells us what the sum of all the field theory
potentials is because there is a dual description in terms of
supergravity. The duality is defined in terms of modular functions
where schematically \eqn\modular{ \sum_{n=1}^{\infty} e^{-nt} =
f(e^{-\pi t}) = f(e^{-\pi\over t}).} Where $n$ is the level of
string excitation defined by \eqn\mass{M_n^2 = ({L\over
\alpha'})^2 + {n+\theta \over \alpha'}} (see Appendix C). For a
parallel D4-brane and D6-brane separated by a distance much larger
than the string scale $M= M_s^2\r$ where $\r >> \ls$, one can
show, by plugging the mass formula \mass\ into the
Coleman-Weinberg formula \vac\ and using a modular transformation
\modular, that the sum \stPot\ is in this case equal to
\eqn\powerLaw{V = M_s^4(\cpl + {M_s \over M}).}}
\def\hwex{

Note that there are several limits one can take in this
Hanany-Witten brane construction. The relevant distances are the
separation of the NS5 branes in the $x^6$ direction, $\l$, and the
distance between the D4-branes and the D6-branes, $\r$, in the 45
direction. If the ratio ${\r\over\l} \ra 0$ and ${\r\over\ls} \ra
\infty$ then we are in the "close supergravity" limit for the 
messenger fields. Here it is
impossible to include massive string states from the 4-6 strings
without including so many 5d Kaluza-Klein states that the theory
on the D4-brane is essentially five dimensional. This is because
in this limit \eqn\closeSUGRA{\eqalign{M_{KK} = {1\over\l} & <<
{1\over \r} \cr M_{mess} = {\r\over \alpha'} & >> {1\over
\r} .\cr}} Since $M_{mess} >> M_{KK}$, to keep even the
lightest string state we have also to keep many Kaluza-Klein
states which propagate in the loop integrals \foot{In this set up
actually Kaluza-Klein modes in the loop integrals cancel to lowest
order due to the fact that they are in $N=4$ multiplets.}. Notice
that in this limit the D4-D6 system is codimension-2 in 10
dimensions, the potential goes as $\log(\r )$ \foot{Note that this
is the same potential that one obtains from the field theory
calculation \ftpot. We believe this to be a coincidence and expect
that the coefficients do not match.}.

The "far supergravity" limit is relevant for four dimensional
physics. This is the limit given by ${\r\over\l}\ra\infty$ and
${\r\over \ls} \ra \infty$. In this limit the KK states are heavy
compared to the massive string states, and we can cut the theory
off before ${1\over \l}$ and still include many massive string
states. In order to do this we demand at least \eqn\bound{
M_{mess} = {\r\over \alpha'}={1\over \l} = M_{KK}} which implies
that $\l << \ls$ as well as 
\eqn\farSUGRA{\eqalign{M_{KK}
= {1\over\l} &
>> {1\over \r}  \cr M_{mess} = {\r\over \alpha'} & >> {1\over \r} .\cr}}
Notice in this limit the D4-brane looks like a 3-brane since $\l$
is small so the $D4-D6$ system is essentially codimension-3 in
10d. We saw in section 2.4 using SUGRA that this potential
indeed goes like ${1\over \r}$. The field theory limit for 
the messenger fields is when
distance $\r$ is small compared to $\ls $ but from the bound
\bound\ which says that the lightest KK mode is greater than or
equal to the lightest messenger field, we find $\l
>> \ls$ .}

\def\HWquantum{
Now let us calculate quantum corrections to this model using tree
level supergravity. In our $N=2$ SQCD set-up discussed above with
non-zero D-term, $\eta \neq 0$, and non-zero adjoint scalar vacuum
expectation value, $<\phi
> \neq 0$, the \CW\ potential can be calculated
using channel duality, as explained in section 2.4, and turns
out to be, \eqn\hwpot{V = { -{\gg^4 \eta^2\ls\over \phi}}.}\hwpot\
says that the adjoint scalar field $\phi$ is no longer flat but
rather has a minimum at the origin. In the ten dimensional
space-time, the interpretation of this potential is that it forces
the D4-brane to roll towards the D6-brane in the 45 direction. If
we are in the limit where $\l << \ls$, the D4-brane can be so
close to the D6-brane that supergravity is no longer a good
approximation; we should use field theory when the lightest mass
state due to the 4-6 strings becomes tachyonic. The theory rolls
to a supersymmetric Higgs branch when the tachyon condenses
\braneReview. Thus the branes seek out the lowest energy
configuration which is the supersymmetric one having zero vacuum
energy.}
\def\twoloop{


We can consider now turning off the Yukawa coupling in the
superpotential in \action\ by setting $\lambda $ to zero.
This breaks the global R-symmetry $SU(1)_J\times SU(2)_R$ 
to $U(1)_J\times U(1)_R$. This
leaves us with $N=1$ SQCD with an adjoint hypermultiplet softly
broken to $N=0$ by the FI-term. In the limit where the FI-term is 
string scale, the global symmetry in enhanced to 
$U(1)\times SU(2)$. 
 The configuration of branes to
which this corresponds is NS 5 in directions 012345, $N_c$ D4 in
directions 01236, and $N_f$ D6-branes in directions 0123456. The
D4-branes and D6-branes are separated by a distance \r. When the
mass for the fundamental fields is sufficiently large, the
lightest modes have the spectrum of a $N=2$ super Yang-Mills
classically.

Lets now calculate quantum corrections to the field
theory using gravity. Now because we turned off 
the Yukawa coupling $\lambda$, the
expectation value of the adjoint scalar field $\phi$ is
independent of the mass of the fundamental fields, $Q$, and
therefore at zero-th order the potential for the adjoint scalars
is still flat but it is shifted by the zero point fluctuations.
The calculation is the same as that in section 3.1 but now
equation \hwpot\ has the interpretation as just a 1-loop vacuum energy
correction rather than a potential of a field and is given by \eqn\NonePOT{V_1
= -{g^4_{YM}\eta^2 \over |m|\ls}} where $m = \r M_s^2$ is a
parameter rather than a field. 


There is a 1-loop beta-function for the adjoint fields due to 
integrating out the massive W-bosons in $N=2$
SYM ignoring all the heavy fundamental fields. \eqn\betaFunction{{1\over
g^2_{ren}} = {1\over  \g} - 2N_c \log{\phi^2\over M^2_{cut}}} Because
corrections to the FI-term, $\eta^2 $ in \NonePOT\ depends on the
coupling constant there is a term in the potential \eqn\renPOT{V_1(\phi) =
{\g\eta^2} (1 + {\g\over m\ls(1 - 2N_c\g \log {\phi^2\over
M_{cut}^2})})} Defining $\phi = M_{cut} - \tilde \phi
$ and expanding \renPOT\ about small $\tilde\phi$, we see that
there is a mass term for the adjoint field. Putting the 1-loop effect of the 
vacuum renormalization together with the 1-loop beta function, we get the 
2-loop diagram shown in figure 7 which corresponds to the order $g^6_{YM}$
effect in the potential \renPOT.


\ifig\twoloopfield{The adjoint scalar acquires a mass from integrating out 
the W-bosons. The W-bosons know about supersymmetry breaking because they couple to 
the heavy messenger fields. This is traditional gauge mediated supersymmetry breaking.
}
{\epsfxsize1.0in\epsfbox{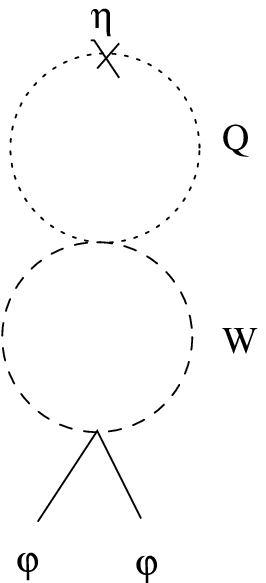}}

\ifig\betaFunc{D4-branes pull on the NS5-branes bending them
in the 6-direction. This is the 1-loop beta function
for the field theory on the branes.}
{\epsfxsize2.0in\epsfbox{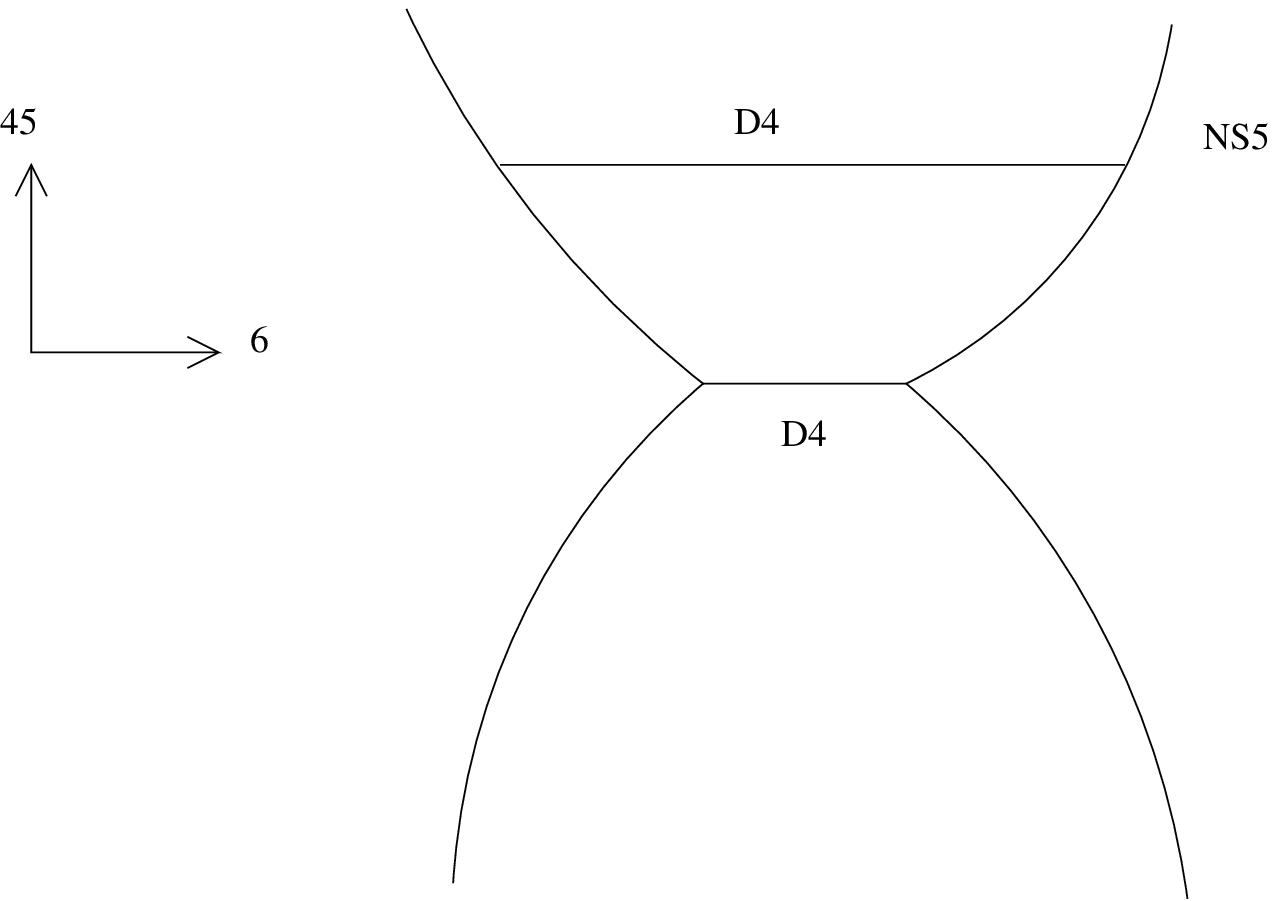}}

In the brane theory, the effective potential \renPOT\ has the 
following interpretation:
the 1-loop beta function corresponds to bending of the NS 5-branes
in the $x^6$ direction due to the D4-branes pulling on them.
However, the D4-brane is pulled in the $x^7$ direction towards the
D6-brane and so the NS 5-branes should bend in that direction as
well. In order for the D4-branes to move in the 45 direction, it
must also move in the 7-direction since it is constrained to  move
along the NS 5-branes. It costs energy for the D4-brane to move
away from the graviational/dilatonic attraction of the D6-branes.
We conclude from this that the bending of the NS 5-branes in the
7-direction corresponds to a 2-loop gauge mediated mass term for 
the adjoint scalar fields. In support of this claim, we point out 
that the denominator of \renPOT\ is effectively a renormalization of 
the mass of the messenger fields as a function of $\phi$ due
to integrating out the massive W-bosons: we can see from 
figure 7 that the loop part of the diagrams for the 
mass renormaliztion of the messengers and the renormaliztion 
of the gauge field coupling constant are exactly the same.
 Indeed we 
see in the brane construction that due to the bending of the 
NS 5-brane in the 7-direction, the length of the stretched strings 
is a function of the 45-coordinates.
Moreover, since the bending of
the NS 5-brane is a 1-loop effect and the open 4-6 strings are
another loop, this agrees with qualitatively with the field
theory.

\ifig\mass{The D4-branes are pulled towards the D6-brane. The
NS5-brane in bent in the 7-direction due to the D4-brane pulling.
The small D4-brane is out of equilibrium and wants to move along
the NS5-brane towards the large cluster of D4-branes. This
corresponds to a 2-loop mass for the adjoint scalar field in field theory
on the brane: one loop is the force between the D4 and D6-branes and 
one loop is due to the brane bending.} {\epsfxsize2.0in\epsfbox{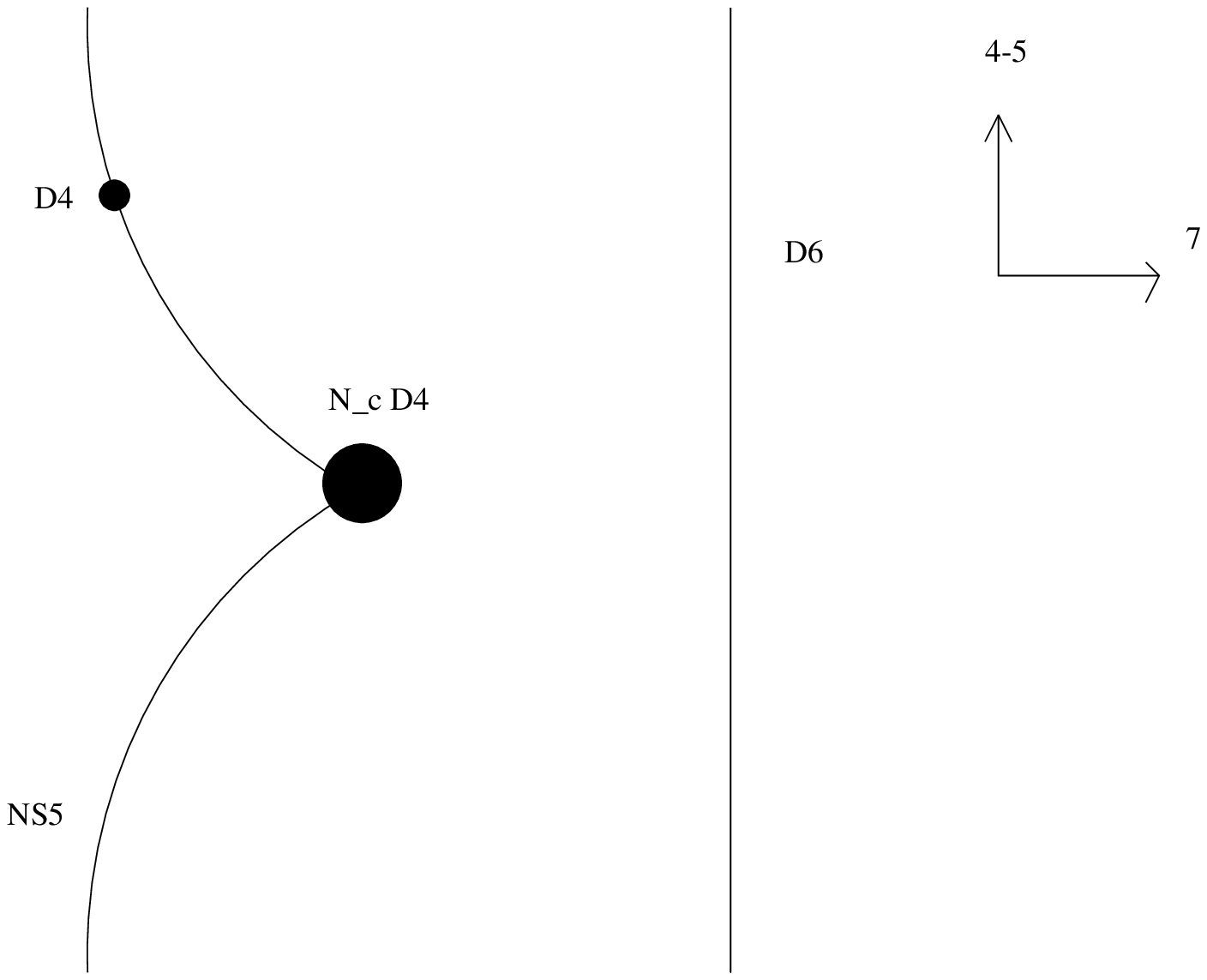}}

Another 1-loop effect is the tadpole correction to the $\eta D$ coupling
at order $\gg^4$ where $D$ is the auxiliary field of $N=1$
superspace. In the brane theory, this has the interpretation as
the bending of the D4-brane in the 7-direction, towards the
D6-brane. Note that the tension of the D4-brane goes like ${1\over
g_s}$ where as the tension of the NS 5-branes goes like ${1\over
g_s^2}$. Therefore, the D4-brane begins to bend at a smaller value
of $g_s $ than the NS 5-brane.

\ifig\bend{The D4-brane bends towards the D6-brane because there
is an attractive force between them. The bending of the D4-brane 
has the interpretation as a 1-loop D-term generation: The vacuum energy shifts but no fields get a mass since the adjoints don't couple to the D-term.
At this order the massless spectrum is still the same as $N=2$ SYM.  Note that
the embedding of the 4-brane here is clearly non-holomorphic
because it depends on on odd number of M-theory coordinates:4567 and 11}
{\epsfxsize2.0in\epsfbox{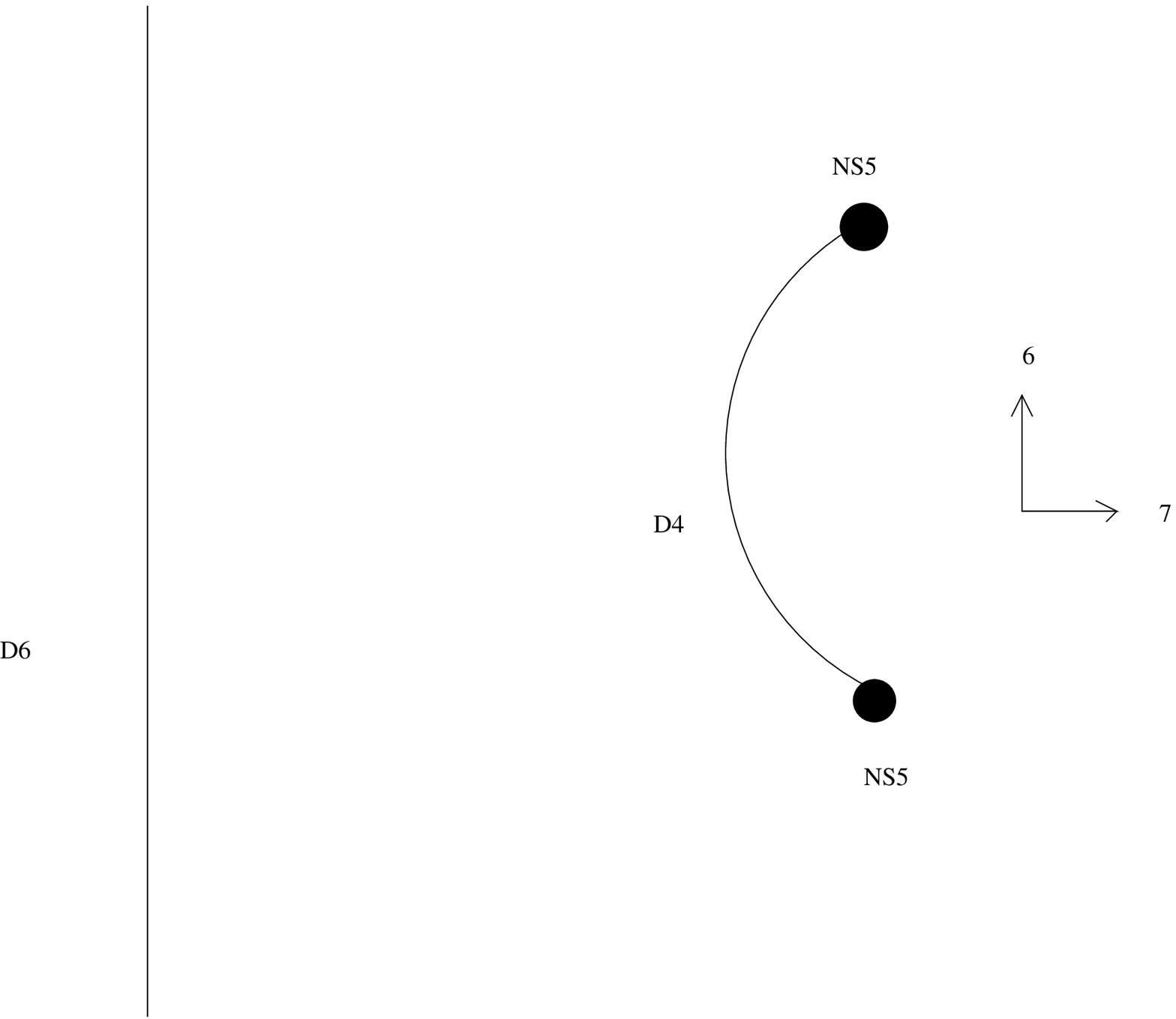}}

Notice that although there is a mass term for $\phi^*
\phi$ coming from \renPOT\and the $SU(2)_R$ symmetry has been broken to a diagonal
$U(1)_R$ there is not a mass term for the fermions. This is
because the fields $\lambda$ and $\psi_{\phi}$ are charged as
$(1,1)$ and $(1,-1)$ respectively under the remaining
$U(1)_J\times U(1)_R$ global symmetry. There is no mass term that
can be generated that respects these symmetries. Even though the
$U(1)_J$ symmetry is broken by instantons to $Z_{4N_c}$ this is
still enough symmetry to prevent a mass term. The low energy
theory is then a \eqn\noSUSY{ {\cal L}  = \cpl F_{\mu\nu}F^{\mu\nu} + \bar\psi
D \psi + \bar \lambda D \lambda}


}
\def\oneloop{

Consider an NS5-brane extending along directions 012345, $N_c$
D4-branes extending along directions 01236, $N_f$ D6-branes
extending along directions 0123689\foot{This configuration is
topologically inequivalent to the one in section 3.1 since there is
a Hanany-Witten transition when the D6-branes cross the NS5.} .
Again the $U(1)_J$ symmetry is broken completely by the mass term
for the hypermultiplets \eqn\massHyper{W = mQ\tilde Q} while the $
SU(2)_R$ global symmetry is broken to $U(1)_R$ explicitly by the
D-term. The D6-brane can move in directions 457 corresponding to 3
real masses. We can turn on the 1 real mass term, $\rho$, which
corresponds to moving the D6-brane away from the D4-brane in the
7-direction. Classically, the $q$s have a mass given by
\eqn\super{V = \sum_n (\sqrt{m^2 + \rho^2} + {n\over \alpha'})
\tilde q_n q_n} where $n$ is the mass level of the 4-6 strings.
Again quantum mechanically the $q$ fields generate a potential for
$\phi$. According to the tree level supergravity, this potential
has the form \eqn\topPot{V = {|\phi |\over \sqrt{\phi^2 + \rho^2}}
(1+{1\over \sqrt{\phi^2 + \rho^2}})} The adjoint scalars have a
mass at $\phi = 0$ when we expand \topPot. At the origin the
$U(1)_J$ symmetry is restored. The potential \topPot\ has a shape
that might be useful in inflationary cosmological senarios. If the
universe rolled along such a potential, then along the flat part
for $|\phi |>> \rho $ the universe would inflate rapidly making it
smooth. Then $|\phi |<< \rho $ is where the universe is now, the
uninflating period.}

\def\higgsB{

In this set up we consider NS 5 012345, NS'5 012389, $N_c$ D4
01236, $N_f$ D6 0123789, and $N'_c$ anti-D4 in directions 01236.
Here it is the D4 and anti-D4, separated in the 45 direction, that
break supersymmetry; the $4-\bar 4$ strings play the role of
non-supersymmetric the messengers rather than the 4-6 strings
which are massless and have a supersymmetric spectrum classically.
The low energy theory on the visible branes is $N=1$ $SU(N_c)$
SQCD with $N_f$ massless flavors $Q$ and $\tilde Q$ as well as
$N_c'$ heavy messenger fields $P$ and $\tilde P$ with mass $\rho =
M_s^2 R$ . Motion of the D4-branes broken on the D6-brane in the
789 direction corresponds to the Higgs branch flat directions of
this $N=1$ theory. We can see for $\g \neq 0$ that the presence of
the anti-D4 branes lifts the Higgs branch flat directions since
there is an attractive force between the D4 and anti-D4 brane that
inhibits the D4-brane from moving in the 789 direction. We
conclude that in the field theory masses for the fundamental
scalars are generated in the same way as in the previous section
for the adjoint scalars. There is a potential similar to equation
\topPot\ but for the $q$ fields as a function of the mass of $P$.
The mass for $P$ explicitly breaks \eqn\HiggsPot{V = {|q |\over
\sqrt{q^2 + \rho^2}} (1+{1\over \sqrt{q^2 + \rho^2}})} the
$U(1)_{45}$ symmetry but not the $U(1)_{89}$ symmetry. Therefore,
the gaugino are protected from getting a mass. The low energy
theory is then one of glue, gauginos, and fundamental fermions.
One can also have chiral fermions using the mechanism discussed in
\brodieHanany\ where the D6-branes split on the NS'5-branes.

There is another phase of the theory when ${\l\over \r}
\rightarrow \infty$ where the D4-brane and the anti-D4-brane
reconnect along the 7-direction. This is then the diagonal
subgroup of the $SU(N_c)\times SU(N'_c)$ gauge theory of the D4
and anti-D4 brane. The force between the D4-brane and the
NS5-brane is now mediated not by strings but by
membranes\foot{Understanding the second order phase transition
here would require "open M-field theory" in analogy with
\senZwiebach.}.}




\newsec{Introduction}

There has recently been a trend 
in particle physics of building models of nature in which
the Standard Model fields lives on a brane while 
gravity is allowed to propagate in more than 3+1 dimensions
\add\rsTwo\rsThree.
Such a scenario is very easily incorporated into string 
theory which naturally has ten dimensions and 
D-branes, walls on which 
open strings propagate \joe.
It is therefore not unfeasible to realize some of the 
large extra-dimension scenarios explicitly in string theory.
However, to do so one would need a mechanism of 
supersymmetry breaking for the brane theories 
since all consistent string theories have
supersymmetry. This is the topic of this paper.
It has been known for some time that it is 
 undesirable to break supersymmetry explicitly 
in the MSSM fields (Minimal Supersymmetric Standard Model) 
 as this generically generates tachyons in 
the visible spectrum\dg. One must first break supersymmetry in some heavy 
non-MSSM messenger fields 
and then communicate the supersymmetry breaking 
to the Standard Model 
fields. Tradiationally the communication was done through the only fields 
which couple to both the MSSM fields and the messenger sector: 
gravity or the Standard Model
gauge fields.
In this paper we will consider all of these same issues in the brane scenario. We find that 
one should not break supersymmetry directly on our brane, instead to avoid having tachyons
one should introduce other branes separted from our brane by extra dimensions.
The role of the messenger fields is played by the strings stretching from 
the visible brane to the hidden brane.
In string theory, because of channel duality which relates properties 
of closed strings to properites of open strings, there are then two ways 
view the supersymmetry breaking mediation. The first way is to view the 
messenger fields is as a tower of extremely heavy open-string 
 fields that couple to the 
massless gauge fields on the visible brane. Integrating out the 
massive messenger fields at 1-loop 
induces masses for the scalars and fermions 
in the visible spectrum. The other way to view the mediation is as 
massless closed string fields which couple to the D-brane world volume 
action giving visible fields masses at tree-level. 
The first method is similar to gauge mediation while the 
second method is similar to gravity mediation. We call our proposal 
``string mediation'' because open-string close-string duality relates 
the two methods.

The mechanism we use for breaking supersymmetry in the messenger open-string sector is 
 rotations in the extra dimensions of the hidden branes with respect to the visible branes.
An important point of our study is that  one can determine if fermion masses
are lifted in a brane construction simply by 
looking at the Lorentz symmetry that is preserved by the 
branes  themselves. These Lorentz symmetries of the extra dimensions 
correspond in the 3+1 brane world to global symmetries 
and can protect the fermions from getting a mass.
It is a somewhat surprising fact that there is a large class of 
non-supersymmetric brane models in string theory which 
have a high degree of global symmetries. 
It is therefore a general results of our study are that many 
generic supersymmetry 
breaking rotations of branes are no good for phenomenology 
because they do not sufficiently break the 
global R-symmetries, generate only a D-term for the 
messenger fields, and therefore do not lift the gaugino.
Fortunately we find that there is  a large class of 
$N=1$ brane constructions where turning on what is often called 
an $F_S$ term is possible and the gaugino can be lifted. 
Scalar potentials are equally easy to understand from the ten 
dimensional point of view: They correspond to 
forces between the branes. Often what will happen in that 
attractive forces between two types of branes will 
be countered by the fact that the branes are stuck to 
larger, heavier branes resulting in a massive scalar in the field theory. 
Although the larger heavier branes 
will bend, the bending is energetically costly and so will balance 
against the attraction of the smaller, lighter branes.
In fact, the bendings themselves often can be interpreted as 
expectation values for D-terms, wave-function renormaliztions,
or other pertubative phenomena.
We find many stable, non-supersymmetric configurations.
One can understand the stability as coming from the 
boundary conditions of the larger, heavier branes 
which extend off to infinity in uncompactifies 
ten dimensional space-time.

The topic of this paper is interesting because eventually one would like 
to model the MSSM using branes in 
string theory (see \antoniadis\fernando)
and one will need to break supersymmetry in a brane model
and calculate the light spectrum of states. String 
mediation seems to be the most natural way to accomplish this.
Moreover, there is some hope that the non-supersymmetric brane 
configurations might be useful for studying 
strong coupling dynamics of non-supersymmetric 
theories.
\foot{ Perhaps as suggested in \brodie\ strong dynamics
can be understood by including only 
Euclidan D0-brane which play the role of instantons in 
the gauge theory but not including Lorentzian D0-branes 
which are five dimensional Kaluza-Klein modes.
This limit is different from the M-theory limit where one 
has both kinds of D0-branes.}

	The outline of the paper is as follows:
In section 2, we review open-closed string 
channel duality and explain the general mechanism of string mediated 
supersymmetry breaking using detached probe branes. In section 3, 
we present some models where the scalars are lifted at 1-loop and 
at 2-loops and a model where the gaugino is lifted at 1-loop.
One of these scalar potentials looks suggestively like an 
inflaton potential that might be relevant for cosmology.
In section 4, we look at the general rotations of branes and explain to 
which spurion each rotation corresponds. In section 5, we explain how 
turning on fluxes is another mechanisms for breaking 
supersymmetry in the messenger sector 
and see how it is related to mechanism of detached rotated probe branes.
 In section 6,
we present a method for having first and second order phase transitions 
in brane constructions which might be relevant for 
modeling the Standard Model Higgs field.

A number of other papers have delt with the issue of supersymmetry breaking 
in brane world scenarios  
such as anomaly mediation \rs, 
gaugino mediation \schmaltz,
radion mediation \luty, 
and in Horava-Witten scenarios \peskin.

\newsec{String Mediated Supersymmetry Breaking}



In this section we
will review open-string closed-string channel duality and explain how it relates to
gauge and gravity mediated supersymmetry breaking on the visible brane.

\subsec{String theory and the annulus diagram}

\annulus

\subsec{Closed string channel and supergravity.}

\sugra

\subsec{Open string channel and 1-loop potentials.}

\potentials

\subsec{Mechanism of tree level 
supersymmetry breaking in the messenger sector: rotation of
               detached probe
brane.}

\probe

\intersection


\newsec{Models of String Mediated Supersymmetry breaking.}

\subsec{Hanany-Witten model with rotated D6-branes: D-term potenial, scaling limits,
Coulomb branch potential at one loop.}

\thicksize=1pt
\vskip12pt
\begintable
\tstrut  |0|1|2|3|4|5|6|7|8|9 \crthick
NS5 |x|x|x|x|x|x|.|.|.|.  \cr
D4  |N|N|N|N|D|D|N|D|D|D   \cr
D6  |N|N|N|N|D|D|D|N|N|N 
\endtable
\noindent

\hwconst

\thicksize=1pt
\vskip12pt
\begintable
\tstrut  |$SO(1,3)_{0123}$|$Spin(2)_{45}$|$Spin(2)_{89}$|$Spin(3)_{789}$|$Spin_{45+89}$ \crthick
$\phi$ |  $\two $| $\hf $ | -$\hf $ | $\two $| 0 \cr
$\lambda_{\alpha}$ | $\two $| $\hf $ | $\hf $ | $\two $| 1 \cr
$\phi$ | 1 | 1 | 0 | 1 | 1 \cr 
q |  1 | 0 | $\hf $ | $\two $| $\hf $ \cr
$\tilde q^*$ | 1 | 0 | -$\hf $ | $\two $| -$\hf $ \cr
$\psi_q$ | $\two $| -$\hf $ | 0 | 1 | -$\hf $ \cr
D | 1 | 0 | 0 | \three | 0 \cr
$F_{\phi}$ | 1 | 0 | 1 | \three | 1 \cr 
$F_{\phi}^*$ | 1 | 0 | -1 | \three | -1 
\endtable
\noindent

\hwex

\trig

\HWquantum

\subsec{Another model with the Coulomb branch lifted at one loop; 
Higgs branch is lifted at tree level.}

\thicksize=1pt
\vskip12pt
\begintable
\tstrut  |0|1|2|3|4|5|6|7|8|9 \crthick
NS5 |x|x|x|x|x|x|.|.|.|.  \cr
D4  |N|N|N|N|D|D|N|D|D|D   \cr
D6  |N|N|N|N|D|D|N|D|N|N 
\endtable
\noindent

\oneloop

\subsec{Higgs branch lifted at one-loop.}

\thicksize=1pt
\vskip12pt
\begintable
\tstrut  |0|1|2|3|4|5|6|7|8|9 \crthick
NS5 |x|x|x|x|x|x|.|.|.|.  \cr
NS'5|x|x|x|x|.|.|.|.|x|x  \cr
D4  |N|N|N|N|D|D|N|D|D|D   \cr
D6  |N|N|N|N|D|D|D|N|N|N   \cr
D4' |N|N|N|N|D|D|N|D|D|D 
\endtable
\noindent

\higgsB

\subsec{Model where gaugino is lifted at one loop.}

Consider NS5-branes along 012345, NS'5 branes along 012389,
D4-branes along 01236, and D6-branes along 0123679 as represented in the
following table. 
\thicksize=1pt
\vskip12pt
\begintable
\tstrut  |0|1|2|3|4|5|6|7|8|9 \crthick
NS5 |x|x|x|x|x|x|.|.|.|.  \cr
NS'5|x|x|x|x|.|.|.|.|x|x  \cr
D4  |N|N|N|N|D|D|N|D|D|D   \cr
D6  |N|N|N|N|D|D|N|N|D|N   
\endtable
\noindent
Let's position
the D6-branes equidistant from the D4-branes in the 48-directions.
The low energy theory on the branes is $N=1$ SQCD with a mass
turned on for the flavors \eqn\massQ{W = mQ\tilde Q} which breaks
the $U(1)_{45}$ symmetry explicitly. Rotation of the D6-brane from
8 into the 6-direction corresponds  to turning on an F-term for m
which explicitly breaks the $U(1)_{89}$ symmetry. Therefore all
global symmetries are broken and there is nothing to protect the
gaugino from getting a mass at 1-loop from coupling to the heavy
messenger fields $Q$. The low energy theory is pure glue. Notice
that there will be non-field theory interactions that are not
visible on the visible brane due to the interaction of the
D6-brane with the NS'5 brane. These interactions are not due to
fundamental strings and so are non-perturbative in string theory.
This appears to be a generic phenomenon when one tries to lift the
gauginos in brane models.


\ifig\gauginoLoop{This is the 1-loop diagram that gives 
a mass to the gauginos. Notice that both a mass term for the messengers 
as well as an F-component must be turned on. Turning on a D-term 
will not lift the gaugino since it won't allow you to draw the diagram in 
this figure. 
An F-term for a mass for a hypermultiplet and a D-term for the vector 
multiplet form a $\bf 3$ of $SU(2)_R$ and are equivalent in 
theories with $N=2$ 
supersymmetry. In $N=1$ theories the $SU(2)_R$ symmetry is broken and 
so the F-term and the D-term are no longer equivalent.}
{\epsfxsize2.0in\epsfbox{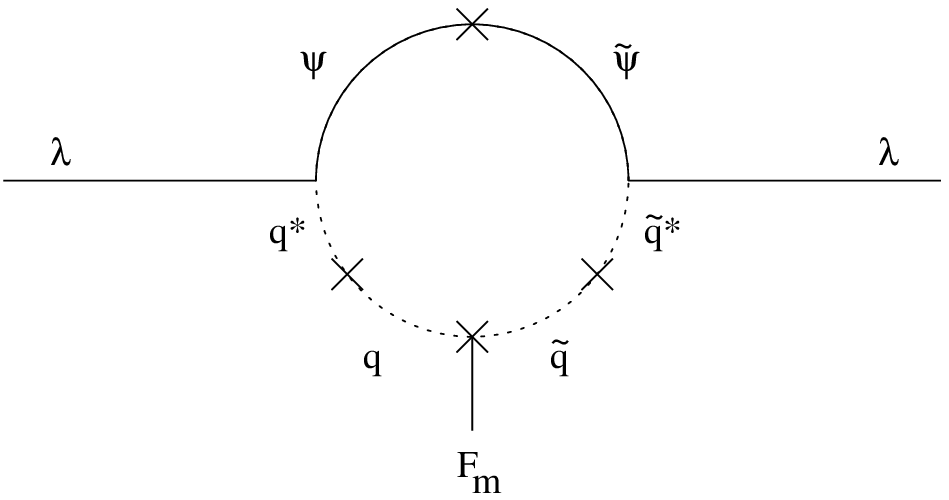}}

\subsec{Two loop mediation.}

\thicksize=1pt
\vskip12pt
\begintable
\tstrut  |0|1|2|3|4|5|6|7|8|9 \crthick
NS5 |x|x|x|x|x|x|.|.|.|.  \cr
D4  |N|N|N|N|D|D|N|D|D|D   \cr
D6  |N|N|N|N|N|N|N|D|D|D 
\endtable
\noindent

\twoloop

\newsec{Soft and Hard breaking from general rotations.}

\subsec{General rotations}

\ifig\hw{These rotations change the Yukawa couplings 
between the adjoints and the fundamentals.}
{\epsfxsize3.0in\epsfbox{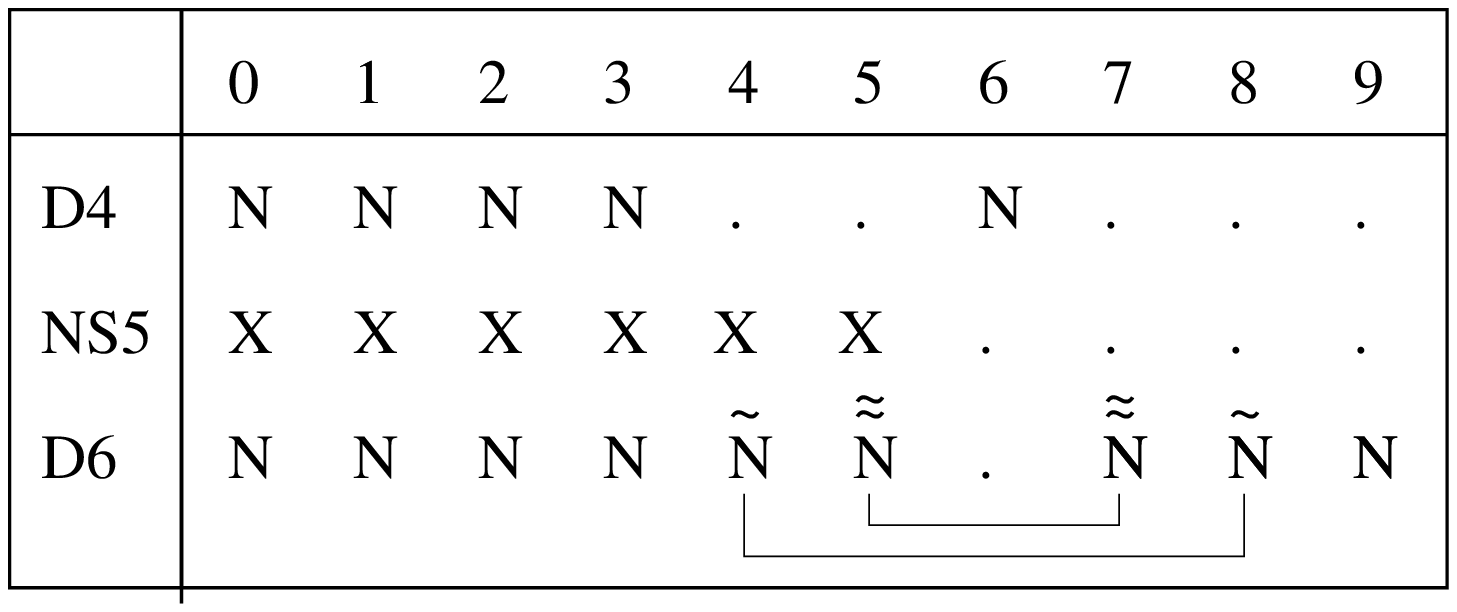}}

\ifig\hw{Both adjoint fermions will have the same mass and rotate 
under $U(1)_{79}$. There is one real massless adjoint scalar
along the 4-direction.}
{\epsfxsize3.0in\epsfbox{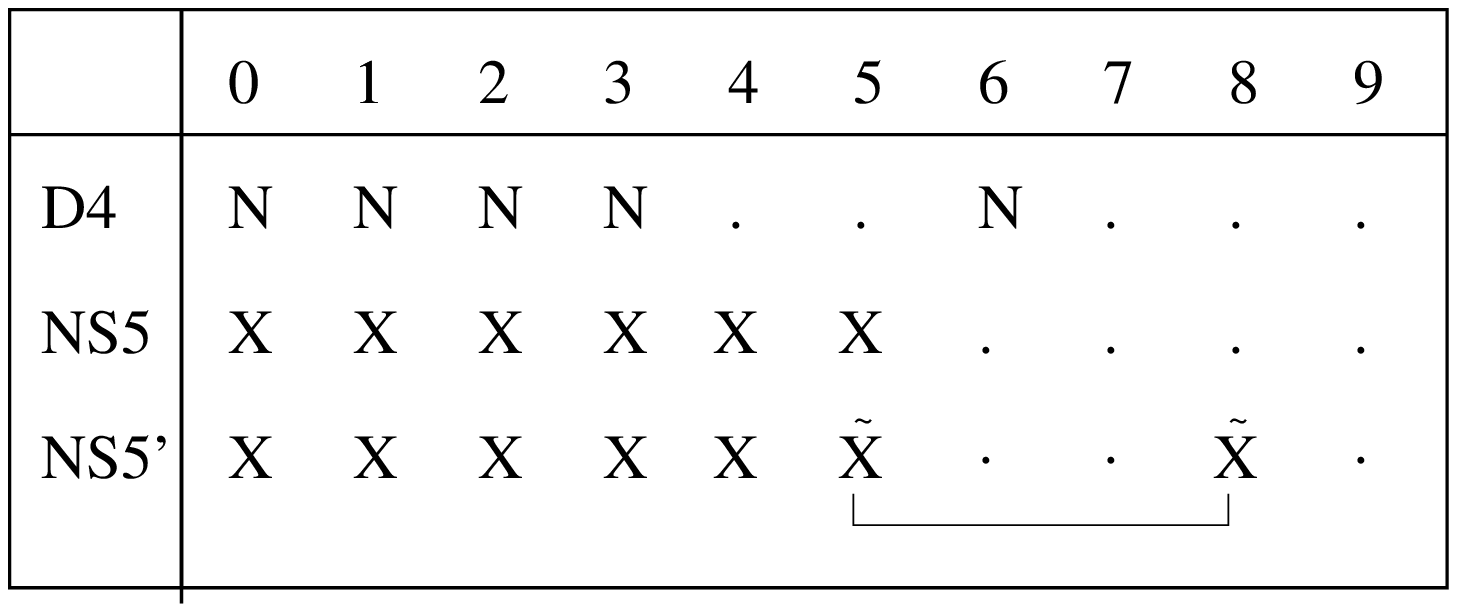}}

\evansParameters

\thicksize=1pt
\vskip12pt
\begintable
\tstrut  |$Spin(2)_{45}$|$Spin(2)_{89}$|$Spin(3)_{789}$|$Spin_{45+89}$ \crthick
$m_Q$ | 1 | 0 | 1 | 1 \cr 
$F_{\phi}$ | 0 | 1 | \three | 1 \cr
D | 0 | 0 | \three | 0 \cr 
$F_{\phi}^*$ | 0 | -1 | \three | -1 \cr
$m_{\phi}$| -1 | 1| \three | 0 \cr
$\zeta$ | -1| 0 | $ \three$ | -1\cr
$\mu $ | -1| -1 | $\three$ | -2 \cr 
$F_{\tau}$ | -1| -1 | $\three$ | -2 \cr
$\lambda$ |  -1 | 1 | $\three$  | 0 \cr
$h^*$ | -1 | 0 | $\three$ | -1 \cr
$\xi$ | -1 | -1 | $\three$ | -2 
\endtable
\noindent

\subsec{The Dimopoulos-Georgi Theorem.} One of the points of this
paper is to show that one cannot just begin with an arbitrary
supersymmetric collection of branes, perform a general rotation,
and hope to obtain a phenomenologically interesting model. The
reason for this is that there will in general be a tachyonic mode
comming from the fundamental strings between the rotated branes in
order that $\str M^2 = 0$ be satisfied before and after the
rotation. Unless this is the Higgs field of the electro-weak
interactions, this would be phenomenologically unrealistic. To get
realistic models, one must use detached probe branes, as we have
been doing in this paper, to break supersymmetry and then mediate
it via messenger fields to the supersymmetric visible sector. This
way one can have light fields in the visible sector that violate
$\str M^2 = 0$. This is not in fact a new result but is the
content of a paper by Dimopoulos and Georgi \dg.

\newsec{Other Mechanisms for Supersymmetry breaking.}

\magnetic

\newsec{First order phase transition and tachyonic instabilities}


Note that there are two limits one can take in the brane
configuration discussed in section 3.1 with NS5 along 012345, D4
along 01236, D6 along 0123456. The NS branes are separated by a
length $\l$ while the D6 and the D4 are separated by a length
$\r$. If length scales are big compared to the string scale $\ls$,
then there is a first order transition. As one decreases ${\r
\over \l}$, the branes will jump from the configuration where the
D4-branes are stretched between the NS5 branes to another more
energetically favorable configuration where the D4-branes split
into D4 and anti-D4 branes stretched between the NS5s and the D6.
A tachyon is never induced during this transition because the
fundamenal strings never become short enough. The phase transition
is from Coulomb to Higgs phase, but instead of having a since
massive W-boson in the Higgs phase there are a whole tower of
closely spaced W-bosons! This is another reason that the "near
supergravity" limit is phenomenologically less interesting than
the "far supergravity" limit. However, if $\l$ is small compared
to the string scale, then as we reduce ${\r \over \l}$, stretched
strings will become so light that a tachyon is induced. There is
then a second order phase transition from a Coulomb phase into a
Higgsed phase. In this limit there will be one W-boson that will
have a much lower mass than the tower of states starting at the
string scale. To understand this transition one might study open
string field theory in analogy with \senZwiebach

\newsec{Lessons}

It is interesting to note that in the brane models that we
considered in section 3 at very short distances in the ten
dimensional space-time each brane looks as though it preserves 16
supersymmetries. It is only when one considers all the branes
together that supersymmetry is broken. This means that the
fundamental theory is in some sense $N=4$ and one can turn on soft
as well as hard breaking terms (from the point of view of $N=1$) 
to reduce it to $N=0$ theories. One
lesson from the brane models is that there may be more parameters
in the MSSM than was previously considered.

Another lesson from the brane models is that certain perverse
couplings from the field  theory point of view look very natural
when one considers branes. For example, consider flavor D-terms.
In the branes one could take a Hanany-Witten model such as the
ones in section 3 with D4-branes, NS5-branes, and $N_f$
D6-branes and choose arbitrary angles $\theta_i$ where
$i=1,...,N_f$ for each of the $N_f$ D6-branes relative to the
D4-brane. In the field theory the $N_f$ angles corresponds to
having $N_f$ D-terms for each of the flavors. It is a strange
operator to consider in field theory, but in the branes it is very
natural.



\newsec{Appendix A}
\subsec{Canonical and string Normalization}

Here we give a map in going from the ``string normalization'' of
the field theory action in equation \action\ and the standard Wess
and Baggar normalization \wb. The Wess and Baggar action is
\eqn\actionWB{\eqalign{{\cal L_{N=2}} & = \int d^2\theta
d^2\bar\theta (\Phi_a^{\dag} e^{\gg V_b} \Phi_c f^{abc} +
Q_i^{\dag} e^{\gg V_a T^a_{ij}} Q_j + \tilde Q_i^{\dag} e^{- \gg
V_a T^a_{ij}} \tilde Q_j + \eta \Tr V) \cr & + \int d^2\theta
W^{\alpha}_a W^a_{\alpha} + \int d^2\theta (mQ\tilde Q+ \lambda
\Phi Q\tilde Q + \Phi [A,B]) \cr }} where the string-normalization parameters are related to the
Wess and Baggar normalization parameters by \eqn\normal{\eqalign{V &
\rightarrow \gg V \cr W_{\alpha } & \ra \gg W_{\alpha } \cr
\eta & \ra {\eta\over \gg} \cr
\Phi & \ra \gg \Phi \cr \lambda & \ra {\lambda\over g_{YM}} \cr }}
The arrow points in the direction of the Wess and Baggar
normalization. In the WB normalization, one can see from the
action \actionWB\ that the mass of the scalar components of $Q$
and $\tilde Q$ is \eqn\massSplit{\eqalign{m^2 & + \gg\eta \cr m^2
& - \gg\eta \cr }} and clearly we recover the supersymmetric limit
when $\gg \ra 0$.

\newsec{Appendix B}

\subsec{One color and one flavor.}

Here we give rules for counting the number of degrees of freedom
on the Higgs branch moduli space using branes. Let us being with
an $N=2$ $U(1)$ theory with $N_f = 1$. There is one vector
multiplet with 8 components and one hypermultiplet with 8
components. Although there is a gauge invariant meson $Q\tilde Q$,
one cannot give it a vacuum expectation value supersymmetrically
because of the D and F-term requirements. We can however turn on a
D-term. This gives a vev to the $q$ and breaks the $U(1)$ gauge
theory. The massive gauge boson eats the massless scalar. In terms
of multiplets: the vector eats the hyper leaving one long
multiplet with 16 components. There are no massless degrees of
freedom after the Higgsing; the massive long multiplet is a stable
non-BPS state since there is nothing for it to decay to.

What is this in terms of branes? This is a simple Hanany-Witten
construction with 2 NS 5-branes, one D4-brane, and one D6-brane
(see section 3.1 ). The 4-4 strings give the 8 components of the
vector multiplet and the 4-6 strings give the 8 components of the
hypermultiplet. Turning on the D-term is like moving one of the NS
5-branes in 789 (or equivalently, rotating the D6 relative to the
D4.) The D4-brane splits into two pieces along the D6-brane
separated in the 789 direction. A string can stretch from one
D4-brane to the other. We identify that string with the 16
components of the massive non-BPS vector multiplet. In no sense
are the ends of the strings charged since the gauge group is
completely broken. Although it appears that there are also 4-6
strings, they must carry no massless states. This is consistent
with the fact that the D4-branes cannot move once the NS5-branes
separate in the 789 direction. To leading order in $g_s$ and
$\alpha'$ this is a state non-BPS brane configuration.

 \subsec{One color and two flavors.}

Now let us consider a $N=2$ theory with a $U(1)$ vector multiplet
(8 components) and $N_f=2$ hypermultiplets (8x2=16 components).
Turning on the D-term gives a mass to the vector multiplet eating
one of the hypermutliplets. However there are still 8 massless
degrees of freedom coming from the uneaten hypermultiplet. Now we
can consider turning on the meson as well in a supersymmetric
fashion.

In the brane construction we have  2 NS 5-branes, 1 D4-brane and 2
D6-branes. We separate the NS 5-branes as before in the 789
direction. This is the D-term. We can also now however move the
middle piece of the  D4-brane between the 2 D6-branes. These are
the massless scalar part of the uneaten hypermultiplet. Now the
massive non-BPS vector multiplet is not stable since it can decay
into the BPS massless hyper.

\newsec{Appendix C}

\subsec{Open string spectrum between two D-branes.}

\spectrum

\newsec{Acknowledgements}
We would like to thank the Aspen Center for Physics, the
Institute for Advanced Studies, Brown University, M.I.T.,
and the Tata Institute for
their hospitality while this work was being completed. We would
also like to acknowledge S. Thomas for collaboration 
on this work as well as N. Evans, C. Johnson, S. Kachru, E.
Gimon, K. Intriligator, D. Lowe, S. Ramgoolam, M. Schmaltz, A.
Sen, M. Strassler, W. Taylor, S. Trivedi, and B. Zwiebach for useful
discussions. We are supported by the D.O.E. under contract number
DE-AC03-76SF00515.

\listrefs

\end